# Multispacecraft Observations of the 2024 September 9 Backside Solar Eruption that Resulted in a Sustained Gamma Ray Emission Event

Nat Gopalswamy[1], Pertti Mäkelä[1,2], Sachiko Akiyama[1,2], Hong Xie[1,2], and Seiji Yashiro[1,2], Stuart D. Bale[3], Robert F. Wimmer-Schweingruber[4], Patrick Kühl[4], and Säm Krucker[3,5]

[1]NASA Goddard Space Flight Center, Greenbelt, Maryland

[2]The Catholic University of America, Washington, DC

[3]University of California at Berkeley, Berkeley, CA

[4] Kiel University, Kiel, Germany

[5]University of Applied Sciences and Arts Northwestern Switzerland (FHNW), Switzerland

ABSTRACT

We report on the 2024 September 9 sustained gamma-ray emission (SGRE) event observed by the Large Area Telescope (LAT) onboard the Fermi satellite. The event was associated with a backside solar eruption observed by multiple spacecraft such as the Solar and Heliospheric Observatory (SOHO), Solar Terrestrial Relations Observatory (STEREO), Parker Solar Probe (PSP), Solar Orbiter (SolO), Solar Dynamics Observatory (SDO), Wind, and GOES, and by ground-based radio telescopes. Fermi/LAT observed the SGRE after the EUV wave from the backside eruption crossed the limb to the frontside of the Sun. SolO's Spectrometer Telescope for Imaging X-rays (STIX) imaged an intense (X3.3) flare, which occurred ~41º behind the east limb, from heliographic coordinates S13E131. Forward modeling of the CME flux rope revealed that it impulsively accelerated (3.54 km $s^{-2}$) to attain a peak speed of 2162 km $s^{-1}$. SolO's energetic particle detectors (EPD) observed protons up to ~1 GeV from the extended shock and electrons that produced a complex type II burst and possibly type III bursts. The durations of SGRE and type II burst are consistent with the linear relation between these quantities obtained from longer duration (>3 hours) SGRE events. All these observations are consistent with an extended shock surrounding the CME flux rope, which is the likely source of high-energy protons required for the SGRE event. We compare this event with six other BTL SGRE eruptions and find that they are all consistent with energetic shock-driving CMEs. We also find a significant east-west asymmetry (3:1) in the BTL source locations.

## 1. Introduction

Solar eruptions occurring far behind the limb (BTL) resulting in >100 MeV gamma-rays observed on the frontside of the Sun have become critical players in identifying the source of high-energy (>300 MeV) protons  responsible for the gamma-rays (Pesce-Rollins *et al.*, 2015a,b; Plotnikov, Rouillard, and Share, 2017; Ackermann *et al.*, 2017; Jin *et al.*, 2018; Gopalswamy et al. 2020; Kouloumvakos et al. 2020; Wu et al. 2021; Pesce-Rollins et al. 2022). Flare reconnection has been suggested as an alternative mechanism for the high-energy particles (Ryan and Lee 1991; Kanbach et al. 1993; Akimov et al. 1996; Grechnev et al. 2018; de Nolfo et al. 2019). However, solar flares typically occupy an angular extent <20º (Yashiro et al. 2013), so it is highly unlikely that protons from the flare site can precipitate on the frontside to cause the sustained gamma-ray emission (SGRE, Share et al. 2017) from the Sun. Pesce-Rollins et al. (2022) reported on the 2021 July 17 BTL event, whose flare properties were obtained from Solar Orbiter's (SolO; Müller et al. 2020) Spectrometer Telescope for Imaging X-rays (STIX; Krucker et al. 2020). These authors concluded that when the EUV coronal wave from the backside eruption reached the visible disk, the >100 MeV gamma-ray emission started, indicating that the coronal wave and the solar energetic particles (SEPs) underlying the SGRE are coupled. They revisited the three previous BTL eruptions associated with gamma-rays to firm up their conclusions.

Forrest et al. (1985) who discovered SGRE had already suggested that the gamma-rays "may be a signature of the acceleration process which produces solar energetic particles (SEPs) in space." The acceleration process for SEPs has been shown to operate in the front of shocks driven by coronal mass ejections  (CMEs; Kahler et al. 1978; Gosling 1993; Reames 1999; 2013; Vainio 2008). The requirement for producing SGRE is the production of >300 MeV protons, which can backpropagate to the Sun from the shock downstream to produce the neutral pions that decay into gamma-rays. Therefore, demonstration of a hard spectrum SEP event in association with the SGRE event ensures the availability of >300 MeV protons.

One of the key aspects of CMEs associated with SGRE is that the CME properties are similar to those with ground level enhancement (GLE) in SEP events. This ensures that >300 MeV protons needed for the production of >100 MeV gamma-rays are present in these eruptions. As in GLE CMEs (Gopalswamy et al. 2016), SGRE CMEs have initial acceleration (> 1 km $s^{-2}$) and are very fast (~2000 km $s^{-1}$), which indicate a hard-spectrum SEP event implying the presence of high-energy SEPs for SGRE production (Gopalswamy et al. 2021; Mäkelä et al. 2024.  In this paper, we focus on the early kinematics of a new BTL event that occurred on 2024 September 9, to show that the underlying CME has impulsive acceleration (>2 km $s^{-2}$) and was very fast (>2000 km $s^{-1}$) resulting in the acceleration of high-energy particles. Metric and interplanetary type II radio bursts were observed by ground-based radio observatories and the FIELDS experiment (Bale et al. 2016) onboard Parker Solar Probe (PSP, Fox et al. 2016). This indicates that particles accelerated early in the event continued to be accelerated throughout the event.

## 2. Observations and Analysis

The 2024 September 9 SGRE event was observed by the Large Area Telescope (LAT, Atwood et al. 2009) on board the Fermi satellite. The SGRE event is listed in the Fermi/LAT quick-look web site to have occurred between 05:52 UT and 06:25 UT with a peak flux of ~$8.1\times 10^{-6}$ photons $cm^{-2}$ $s^{-1}$ (https://hesperia.gsfc.nasa.gov/fermi/lat/qlook/lat_events.txt). The Fermi satellites day began at 05:39 UT and ended at 06:38 UT, while the Sun exposure was between 05:50 and 06:28 UT. The associated CME was well observed by SOHO's Large Angle and Spectrometric Coronagraphs (LASCO, Brueckner et al. 1995) C2 and C3 and by STEREO's Sun Earth Connection Coronal and Heliospheric Investigation (SECCHI, Howard et al. 2008) coronagraphs (COR1, COR2) and the Extreme Ultraviolet Imager (EUVI). The hard and soft X-ray flare observations from the Spectrometer Telescope for Imaging X-rays (STIX) (Krucker et al. 2020) on SolO is the only instrument with direct observations of the solar source providing time profile and flare images. The eruption was also observed in EUV by SDO's Atmospheric Imaging Assembly (AIA, Lemen et al. 2012) and STEREO/SECCHI's Extreme Ultraviolet Imager (EUVI). A long-duration metric type II burst was observed by several ground-based observatories including the CALLISTO (Compound Astronomical Low frequency Low cost Instrument for Spectroscopy and Transportable Observatory, Benz et al. 2005) network and the Radio Solar Telescope Network (RSTN). Radio emissions were recorded by the Radio and Plasma Wave experiment (WAVES) on board Wind and STEREO (Bougeret et al. 1995; 2008), SolO's Radio and plasma wave (RPW, Maksimovic et al. 2020) experiment and PSP's FIELDS instrument. An SEP event was reported by NOAA's GOES satellite, but the origin of the SEP event is ambiguous because there was a west-limb CME around the onset time of the SEP event. Fortunately, SolO's energetic particle detector (EPD, Rodriguez-Pacheco et al. 2020) fully observed the SEP event. These observations help us develop a comprehensive picture of the SGRE event, with all required features covered: an ultrafast CME with impulsive acceleration, an X3.3 X-ray flare, a type II burst extending into the interplanetary domain, and a large SEP event with hard spectrum.

The relative positions of the spacecraft observing the 2024 September 9 SGRE event are shown in Fig. 1. Based on the SolO/STIX observations, we infer that the eruption location to be S13E131 (heliographic coordinates). At the time of the eruption, SolO was located at a heliocentric distance of 0.53 au at S04W172.9, almost directly behind Earth. Thus, the eruption location is magnetically well connected to SolO. The onset of the event is marked by the increase in SolO/STIX soft X-ray emission. The CME had its nose in the northeast direction that eventually become a halo in SOHO/LASCO and STA/COR2 field of view (FOV). The CME had a sky-plane speed of ~1533 km $s^{-1}$ in the LASCO FOV. The overall timeline of the event is shown in Table 1.

**Table 1**. Timeline of the 2024 September 9 eruption

| Time (UT)[a] | Event | Mission/instrument |
|---|---|---|
| 04:56 | Soft X-ray flare onset (4-12 keV) | SolO/STIX |
| 04:57 | Hard X-ray onset (32-76 keV) | SolO/STIX |
| 04:58 | CME first appearance in the AIA FOV | SDO/AIA |
| 05:01 | First major hard X-ray peak | SolO/STIX |
| 05:05 | CME first appearance in the EUVI FOV | STA/EUVI |
| 05:07 | Onset of metric type II burst at ~70 MHz | CALLISTO |
| 05:10 | Last hard X-ray peak | SolO/STIX |
| 05:11 | CME first appearance in the COR1 FOV | STA/COR1 |
| 05:12 | Soft X-ray peak (≈ SGRE onset) | SolO/STIX (Fermi/LAT) |
| 05:14 | First peak of DH type III bursts | SolO, PSP, STA, Wind |
| 05:15 | Onset of SEP event at >400 MeV | SolO/EPD-HET |
| 05:18 | DH type II onset (flank and nose) | Same as above |
| 05:22 | Limb crossing of EUV wave from backside | SDO/AIA |
| 05:24 | CME first appearance in LASCO/C2 FOV | SOHO/LASCO |
| 05:33 | End of metric type II burst at 20 MHz | CALLISTO |
| 05:39 | Beginning of Fermi day | Fermi |
| 05:44 | End of DH type III group at 1 MHz | PSP, STA, Wind |
| 05:50 | Beginning of Sun exposure | Fermi/LAT |
| 05:52 | SGRE flux 4.2 sigma above background | Fermi/LAT |
| 06:25 | SGRE flux dropped below 4.2 sigma of background | Fermi/LAT |
| 06:28 | End of Sun exposure | Fermi/LAT |
| 06:30 | End of type II, probably extends to 7:00 UT | PSP |
| 06:38 | End of Fermi day | Fermi |
| 06:57 | End of SGRE (estimate) | Fermi/LAT |

[a]SolO and PSP times shifted to Earth

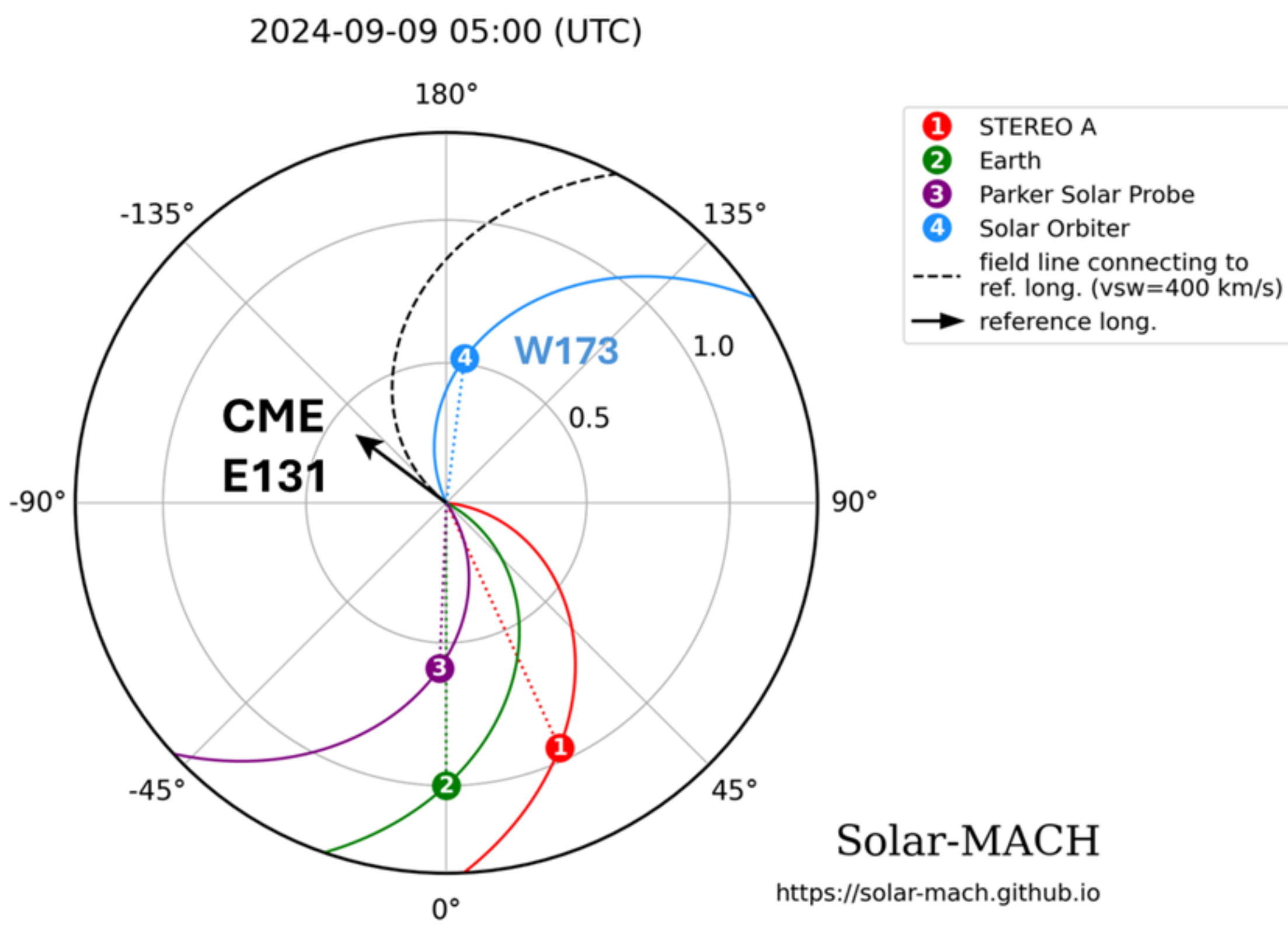

**Figure 1**. The locations of various spacecraft in the inner heliosphere near the onset time of the 2024 September 9 eruption from solar-MACH (Gieseler et al. 2022). Earth represents SOHO, Wind, and ground-based observatories. The arrow points to the direction (E131 in Earth view) of the CME based on the flare location estimated from SolO/STIX located at W173.

### 2.1 X-ray Observations of the Solar Source

SolO was at a heliocentric distance ~0.53 AU at a longitude of W172.8 in Earth view (i.e., directly behind the Sun as seen from Earth). Figure 2a shows the time profiles of thermal (4-12 keV) and non-thermal (32-76 keV) X-ray flux in normalized count flux. The thermal emission (soft X-ray flare) starts around 04:56 UT and peaks at 05:12 UT indicating a rise time of ~16 min. The time profiles are plotted only up to 05:25 UT because at later times, the SEPs accelerated during the event hit the spacecraft producing secondary X-rays that largely enhance the STIX background, especially at high energies.

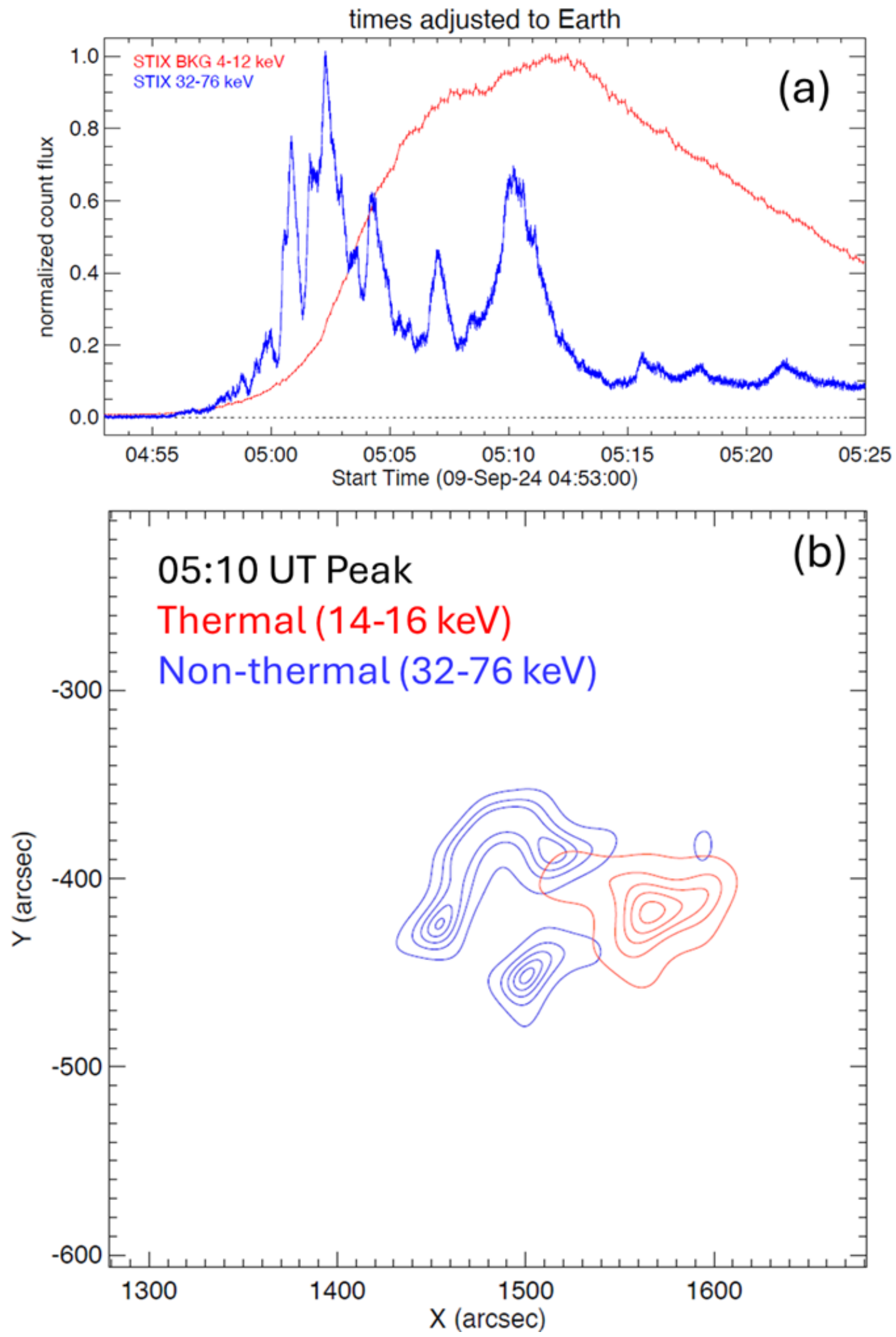

**Figure 2.** (a) Light curves in soft (red, 4-12 keV) and hard (blue, 32-76 keV) X-ray channels. The time has been adjusted to Earth time by adding ~4 minutes to the STIX time. (b) Thermal (red contours) and nonthermal (blue contours) X-ray emission from the eruption. The images have been made with a 20 s integration time. The contour levels are at 10, 30, 50, 70 , 90% of the peak intensity. The eruption is from the southwestern quadrant of the Sun. The thermal loop is to the west as expected from projection effects. The solar radius is ~1811.32 arcsec in SolO view.

The impulsive phase marked by the nonthermal emission (hard X-ray flare) in the energy range 32-76 keV has several nonthermal peaks. The five major peaks are at 05:01, 05:02, 05:04, 05:07, and 05:10 UT and occurred before the thermal peak at 05:12 UT. Nonthermal and thermal images were made during the last peak (05:10 UT) and are shown in Figure 2b. The 32-76 keV contours show the flare ribbon structure with a long (~120 arcsec) northern ribbon. The overall extent of the ribbons in the north-south direction is ~130 arcsec. The source centroid is at (1500", -420"), which correspond to S13W56 in STIX FOV. The flare location thus becomes S13E131 or 41$^{o}$ behind the east limb in Earth view. We also note that the angular size of the flare structure is much smaller than the angular distance to the Earth view. The hard X-ray ribbons and the soft X-ray loop structure connecting the ribbons show that the neutral line is slightly tilted to the horizontal. The source information derived from STIX data is an important input to the forward modeling of the CME flux rope (see section 2.5.2). Based on the method developed by Stiefel et al. (2025), STIX soft X-ray observations give a GOES-equivalent X-ray flare size as X3.3±0.4. The online tool that displays light curves at various channels has the provision to compute the GOES equivalent flare size (https://datacenter.stix.i4ds.net/view/ql/lightcurves).

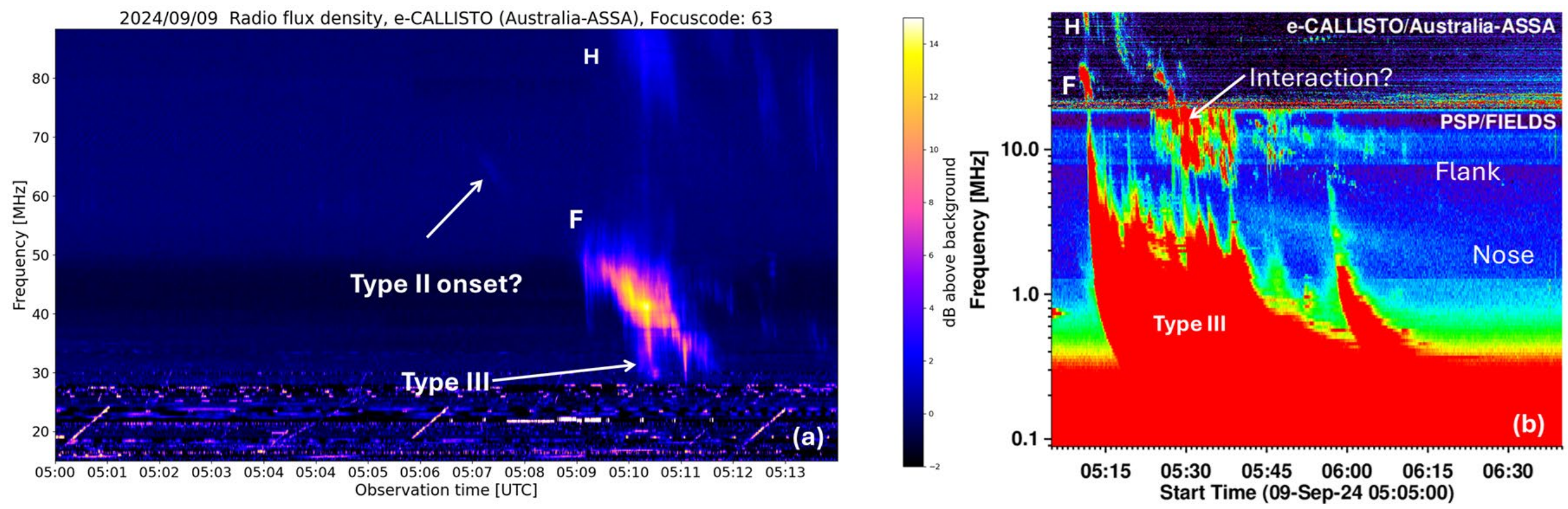

**Figure 3**. (a) Radio dynamic spectrum from the e-CALLISTO station ASSA in Australia (Monstein et al. 2023) showing the onset of metric type II burst with fundamental (F) – harmonic (H) structure. There is a faint extension of the F component to earlier times (05:07 UT) at higher frequency (~67 MHz). A herringbone-like extension can be seen extending from the type II burst to lower frequencies is marked as type III. There is a fainter correspondence in the H component as well. (b) Radio dynamic spectrum from PSP/FIELDS with the e-CALLISTO dynamic spectrum extended beyond (a) and superposed at the top. A complex type III is marked. The type II burst has multiple components, which we identify as “Flank” and “Nose.” The F and H components from (a) are also noted in (b).

## 2.2 Type II and Type III Radio Bursts

The eruption involved type II radio bursts in the metric and decameter-hectometric (DH) domains. The metric type II burst was observed by many stations of the ground-based e-CALLISTO network. Here we use the dynamic spectrum from the Astronomical Society of South Australia (ASSA) station. In the ASSA dynamic spectrum in Fig 3a, one can see that the fundamental (F) structure of the metric type II burst. The F component starts at 5:07 UT with a starting frequency of ~67 MHz. The F component becomes very strong a couple of minutes later (05:09 UT, 50 MHz). The starting frequency of the harmonic (H) component is not seen until ~5:10 UT when it drifts down to ~90 MHz. The H-component lasts until ~5:30 UT and its continuation can be seen as the DH component around this time as observed by PSP/FIELDS and lasting until 06:00 UT. The continuation of the F-component is not clear because of a broadband signature between 05:20 and 05:40 UT that resembles a CME interaction signature (Gopalswamy et al. 2001). The continuation of type II bursts from the metric to DH domain (the so-called m-DH type II burst) is thought to originate from the shock flanks. A second type II feature starts below 10 MHz around 05:30 UT and extends beyond 06:30 UT in the FIELDS dynamic spectrum. The feature is not harmonically related to the m-DH type II burst noted above. This component is thought to originate from the shock nose, which is at a larger heliocentric distance and hence emits at a lower frequency. The nose component often extends to kilometric (km) domain and is also referred to as DH-km type II burst (see Gopalswamy 2011 for other examples).

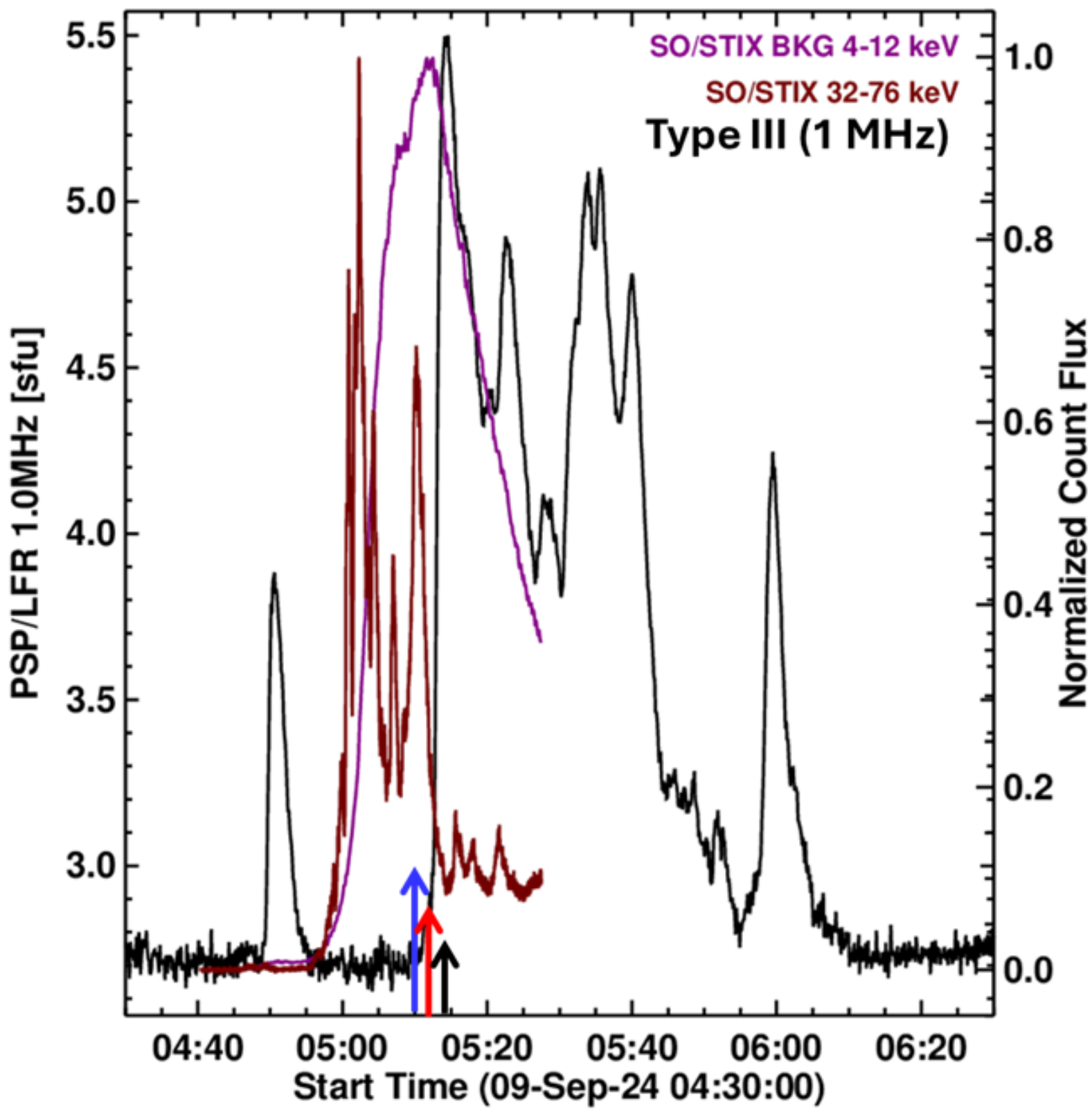

**Figure 4**. Time profile the DH type III burst at 1 MHz (black curve) superposed on the soft (4-12 keV) and hard (32-76 keV) X-ray curves. The last hard X-ray peak occurs at 05:10 UT (blue arrow). The soft X-ray peak is around 05:12 UT (red arrow). The first type III peak occurs just before 05:15 UT (black arrow). The entire type III activity is after the impulsive phase of the flare. The most intense type III burst peaked around the same time in SolO/RPW and PSP/FIELDS dynamic spectra. The bursts are also observed at STEREO and Wind about 4 minutes later as expected given their relative locations. SolO, Wind, and STA radio data are not shown. PSP and SolO are at different longitudes, but roughly at the same heliocentric distance (111.5 Rs vs. 125.9 Rs), so the light travel time is longer at PSP only by 0.6 min.

Some peculiarity of the intense DH type III burst in this event is worth noting. The DH type III burst seems to be formed by the merger of several type III bursts appearing to start from the nose component of the type II, but not all of them. Some seem to originate from the flank component as well. In fact, the first major DH type III burst seems to originate from the fundamental component of the m type II burst (see Fig. 3a). The starting tips of the type III bursts originating from the nose component start at lower frequencies at later times, mimicking the drift of the nose component. Gopalswamy (2011) presented a similar event (2005 August 23) in which the type III-like bursts originated from the fundamental and harmonic components of a DH type II burst. This is an important finding that suggests that not all eruption type III bursts are from the flare site.

Additional support to the shock origin of the DH type III burst is revealed by the delay between hard X-ray and DH type III bursts. Figure 4 compares the time profiles of the STIX X-rays and the FIELDS type III bursts at 1 MHz. Based on PSP/FIELDS and SolO/RPW data, the onset of type III burst activity is only at ~05:10 UT, which is the peak time of the last hard X-ray burst. The first type III peak occurs at least 4 minutes after the last HXR peak. Furthermore, the first type III peak is delayed by ~2 minutes from the SXR peak. Clearly, the onset of the type III burst activity and the subsequent type III peaks occur after the impulsive phase of the X3.3 flare. Recall that the onset of type III activity is also delayed with respect to the metric type II burst observed by e-CALLISTO (see Fig. 3) by several minutes. The delay between type III activity and hard X-ray bursts is in stark contrast to DH type III bursts coinciding with HXR bursts during the impulsive phases of a few flares reported by James and Vilmer (2023). These authors concluded that the electrons responsible for HXR emission and the DH type III bursts are from the same acceleration site. It must be noted that the bursts studied by James and Vilmer (2023) are simple events that do not involve type II bursts.

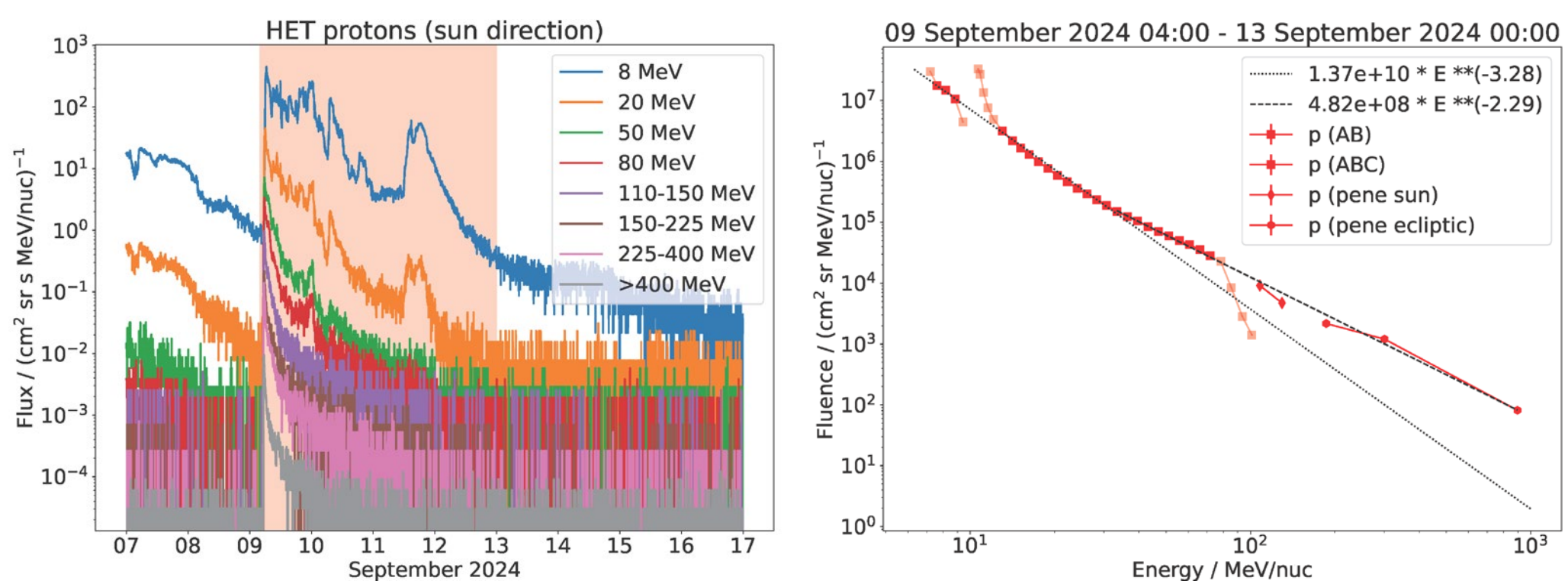

**Figure 5**. Proton flux from SolO/EPD in selected energy channels (left) and the fluence spectrum (right). The highest energy channel is an integral channel recording protons at energies >400 MeV. This channel and the 225 – 400 MeV differential channel are relevant for SGRE emission. Some data points near 10 MeV and 100 MeV are from channels that do not have adequate calibration, so they are not included in the spectral fit. The spectral index for the entire range is $2.29 \pm 3.2\times10^{-3}$. A fit to data points <50 MeV yields a softer spectrum with a spectral index of $3.28 \pm 6.5\times10^{-4}$. The uncertainty in the spectral index is estimated based on the uncertainty in fit coefficients of the regression line.

### 2.3 Solar Energetic Particles

As noted in Fig.1, SolO was well connected to the 2024 September 9 eruption, so the SolO/EPD fully observed the SEP event at energies from keV to GeV. Figure 5 shows the proton flux observed by SolO/EPD at several energy channels of its sunward-pointing telescope. The 225 – 400 MeV differential channel and the >400 MeV integral channel are highly relevant to the production of SGREs. The nose connectivity of SolO/EPD to the eruption site ensures that it does not miss any high-energy particles accelerated at the shock nose. Furthermore, the fluence spectrum in Fig. 5 shows that it is hard with a spectral index of 2.29. The spectral index is similar to that (2.28) of the largest SEP event in solar cycle 24 observed by STA on July 23, 2012 (Russell et al. 2013; Gopalswamy et al. 2016). In fact, the average fluence spectral index of western hemispheric SEP events in the 10-100 MeV energy range is 2.68 when they have a GLE component (Gopalswamy et al. 2016). The hard spectrum of the 2024 September 9 SEP event indicates the presence of ~GeV particles, so >300 MeV particles are readily available.

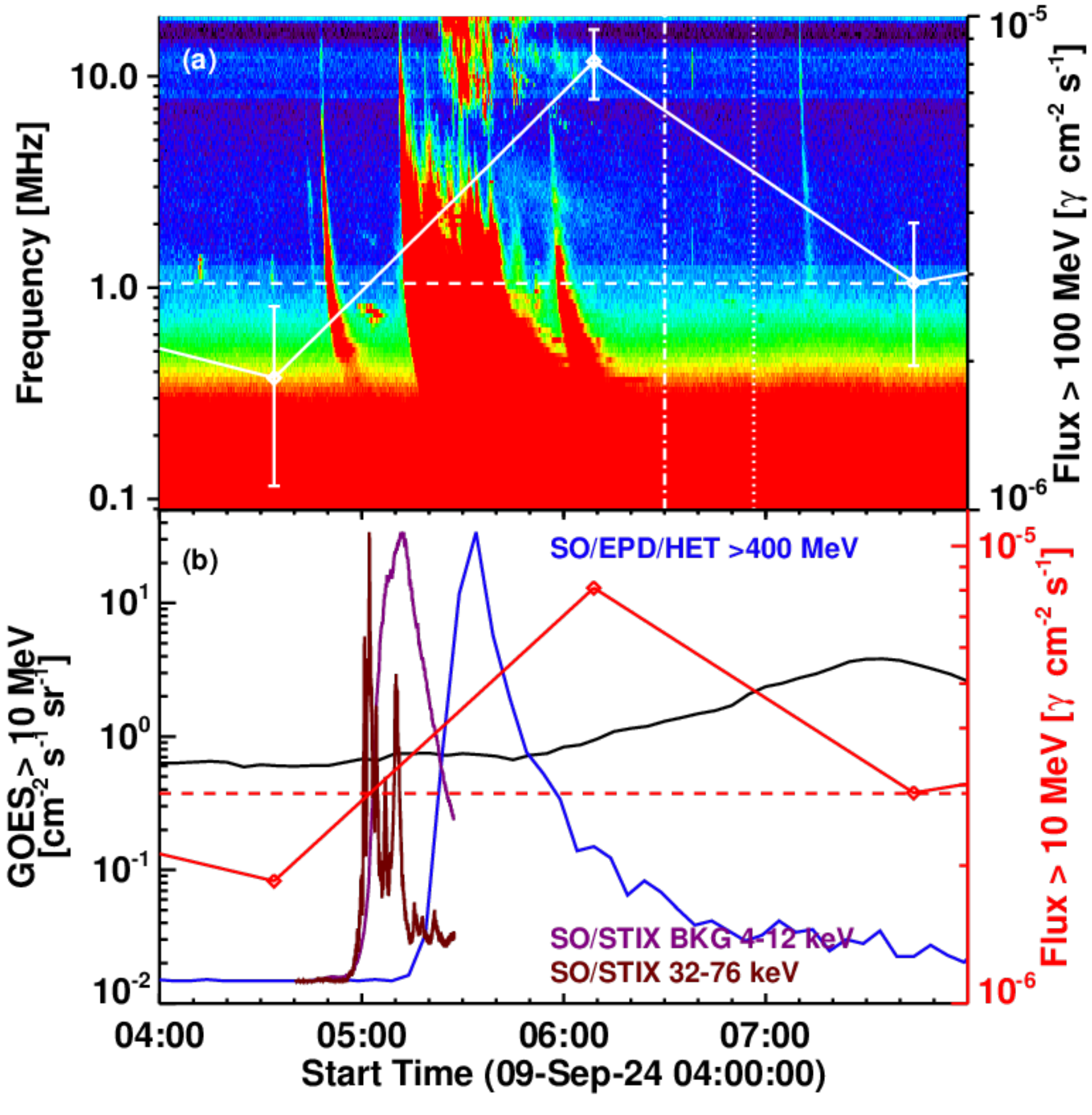

**Figure 6.** (a) Overlay of the SGRE light curve (white curve) on the PSP/FIELDS dynamic spectrum. The ending times of the type II burst and the SGRE event are marked by the vertical dot-dashed and dotted lines, respectively. The three red data points on the SGRE curve correspond to the times of maximum Sun exposure (04:35, 06:09, and 07:44 UT). The 04:35 UT

exposure ended before the soft X-ray flare, so the SGRE must have started somewhere between 04:35 and 06:09 UT. No SGRE was observed during the 07:44 UT exposure. Therefore, the SGRE must have ended sometime between 06:09 UT and 07:44 UT. (b) The SGRE light curve (red curve) compared with STIX HXR (32-76 keV) and SXR (4-12 keV) flare light curves (purple curves) and the >400 MeV protons from SolO/EPD high energy telescope (HET, blue curve). The GOES >10 MeV time profile is also shown (black curve).

### 2.4 The SGRE Event Compared with the Type II burst, X-ray flare, and SEPs

Figure 6 compares the time profiles of various energetic phenomena associated with the SGRE event. The SGRE emission was observed during the 06:09 UT Fermi/LAT Sun exposure (05:50 UT-06:28 UT), well beyond the impulsive phase of the flare as expected for an SGRE event. The previous Sun exposure was between 04:17 and 04:52 UT, before the start of the soft X-ray flare, so we do not know if there was >100 MeV emission in the impulsive phase of the flare. In order to exclude the impulsive phase gamma-rays from SGRE, we take the starting time of the SGRE event as the time of the soft X-ray flare peak (05:12 UT), according to the procedure adapted in previous publications (Gopalswamy et al. 2018; 2019). The existence of a metric type II burst starting at 05:07 UT indicates the presence of a shock accelerating particles. Thus, by 05:12 UT, the SGRE production would have started, but is not observed on the front side. The on-disk SGRE was observed during the 06:09 UT exposure. During the next exposure at 07:44 UT (07:17 UT-08:01 UT) the SGRE was not observed. Therefore, the end time of the SGRE is defined as the mid time (06:57 UT) between the 06:09 UT and 07:44 UT exposures. The total duration of the SGRE is thus 1.75±0.79 hr (05:12 UT to 06:57 UT). The uncertainty is half of the interval 06:09 UT-07:44 UT.

The duration of the DH type II burst extends from 05:18 to 06:30 UT in the FIELDS dynamic spectrum. A close examination of the dynamic spectrum indicates that the type II might have ended around 7:00 UT, indicating a total duration of 1.45 ± 0.25 hr. This is likely to be an underestimate because only a small area of the shock surface is "visible" to FIELDS. The SolO/RPW data are not available, which might give a better estimate of the type II duration. Given that the metric type II burst started about 10 minutes earlier, the onset time is not in large error. Nevertheless, the end times of the SGRE and the type II burst are similar, and the durations are consistent with the linear relation obtained for SGRE events with duration >3 hr (Gopalswamy et al. 2019). The SoloO/EPD/HET proton flux profile in Fig. 6 shows that the SEP event starts around 05:15 UT in the >400 MeV energy channel. The solar particle release time is expected to be a couple of minutes earlier, in agreement with the SGRE start time.

The >400 MeV flux remains elevated past the SGRE event. This is also true for the 225-400 MeV channel, which is also relevant for SGRE production. Given the hard spectrum of the SolO SEP event, we see that the required high-energy particles were accelerated in the eruption. There was a large SEP event observed by GOES (30 pfu) in the >10 MeV channel. However, there was no enhancement in the higher energy channels, possibly due to poor connectivity to a BTL eruption. Furthermore, there was a preceding west limb halo CME around the time of the backside eruption, so it is not clear if we can associate the GOES proton flux enhancement with the backside eruption.

### 2.5 CME Properties

We now describe the CME that was responsible for driving the shock and the consequent SEPs and type II radio burst. The main purpose is to show that the CME is energetic enough to be an SGRE driver. We track the CME manifestation in the FOV of EUV images and coronagraphs. We also provide evidence that the EUV waves associated with the CME propagate on to the solar disk in Earth view to confirm the possibility of energetic particle precipitation on the frontside. After describing observational properties, we show 3-D CME properties using graduated cylindrical shell (GCS) flux rope (Thernisien 2011) fits to multiview (SOHO, STEREO) observations.

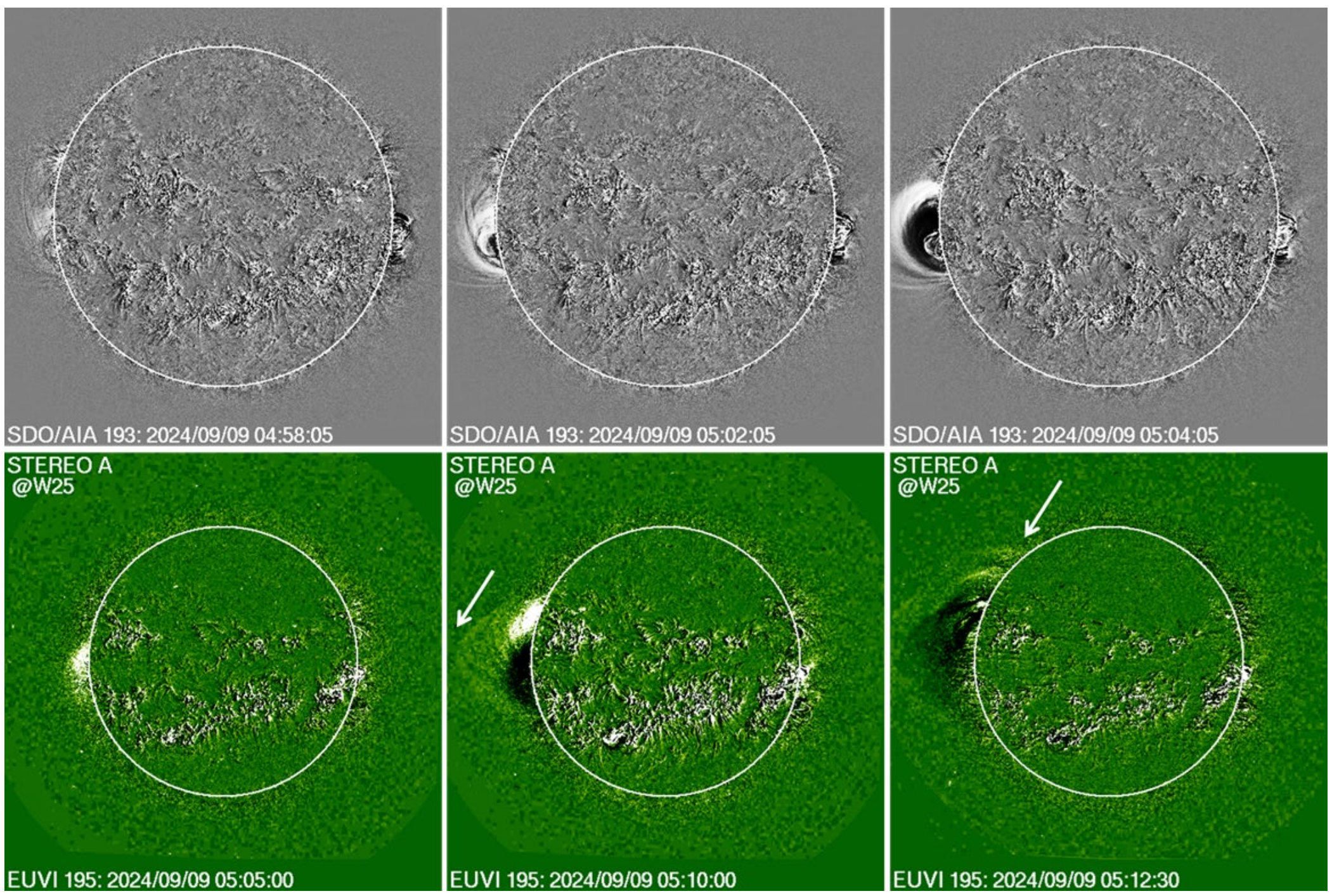

**Figure 7.** Early appearance of the backside eruption in EUV images obtained by SDO/AIA (top) and STA/EUVI (bottom). All are running difference images. At 04:58 UT, the EUV disturbance first appeared in the SDO/AIA FOV as a tiny feature, which occupies a much larger volume in the next 4 minutes. In the 05:04 UT frame, the leading edge (LE) left the SDO/AIA FOV, and there is an intensification of the northern flank. A similar pattern in the STA/EUVI frames but the LE is at larger heights. The first appearance in EUVI is delayed by ~7 min because STA is at W25. The LE and the northern edge are marked by arrows in the 05:10 and 05:12 UT frames.

#### 2.5.1 Early Life of the Backside CME

Figure 7 shows the earliest stages of the backside eruption as observed by SDO/AIA and STA/EUVI instruments, which observed the CME very close to the Sun. In SDO/AIA 193 Å image, the EUV disturbance appears as a small brightness enhancement above the east limb at 04:58 UT. By 05:02 UT, the eruption can be seen as a closed structure with its leading edge (LE) at a height of ~1.25 Rs at a position angle of 92º. Given the eruption location ~41º behind the limb, the LE is likely to be at a height of ~2.2 Rs using a simple deprojection. In the next frame at 05:04 UT, the LE has already crossed the SDO/AIA FOV, but the northern flank is much brighter and propagating faster than the southern one. The STA/EUVI FOV is larger, and the viewpoint is from a larger angle (W25), so the LE can be observed over a longer period of time. At first appearance (05:05 UT), the LE is at 1.18 Rs, which deprojects to 2.9 Rs. Five minutes later, the LE is almost at the edge of the EUVI FOV with a height of 1.58 Rs (3.88 Rs deprojected) at 05:10 UT, when the type III activity started and the last HXR burst peaked. At 05:12 UT, the LE crossed the EUV FOV.

Figure 8 shows the continued evolution of the CME in the FOV of various coronagraphs. At 5:11 UT, about one minute after the EUV CME reached the EUVI FOV, the CME appeared in the COR1 FOV with a LE height of 1.77 Rs in the sky plane at PA = 95º. In the LASCO FOV, the CME appears at 05:24 UT with the LE at 3.7 Rs at PA = 75º indicating a northward deflection of by ~25º. The first appearance in the COR2 FOV is at 05:38 UT with a LE height of 5.4 Rs at PA = 52º, which is ~43º north of initial location. When the backside active region (AR) rotated to the frontside three days later, SDO/AIA image shows a large coronal hole to the south and west of the AR, which might have caused the large deflection.

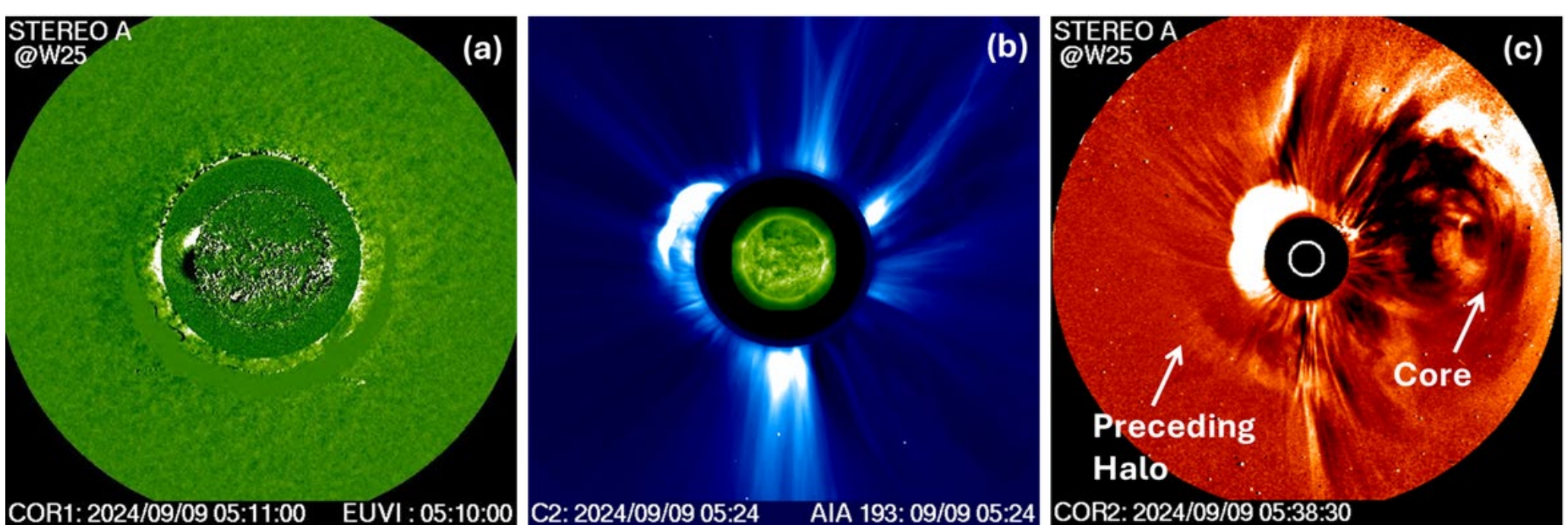

**Figure 8**. The first-appearance frames of the 2024 September 9 CME in the FOV of various coronagraphs: (a) STA/COR1 at 05:11 UT, (b) SOHO/LASCO at 05:24 UT, and (c) STA/COR2 at 05:38 UT. The different first appearance heights are due to the different longitudinal distances of the eruption site with reference to SOHO and STEREO. The arrows in (c) point to the eastern part of the preceding west-limb halo CMEs and its core. The northward deflection of the CME is clear in the LASCO and COR2 frames. The eastern section of the west limb halo that preceded the event in question and its core are marked in (c).

One of the consequences of the northward deflection is that the EUV wave associated with the CME crosses the limb in SDO/AIA at large northern latitudes. The series of running-difference images in Fig. 9 from SDO/AIA at 193 Å shows the northward propagation of the EUV disturbance behind the east limb from the low-latitude region and crossing the limb around 05:22 UT at PA=43º. The disturbance can be seen until 05:34 UT on the disk. In movies, one can see the disturbance until about 05:38 UT beyond which it could not be distinguished from the background. The STA/EUVI difference images do show the EUV disturbance was also discernible in the EUVI images at 05:32 and 05:40 UT. This is a key piece of evidence supporting the possibility of high-energy particles precipitating to the frontside photosphere behind the shock front.

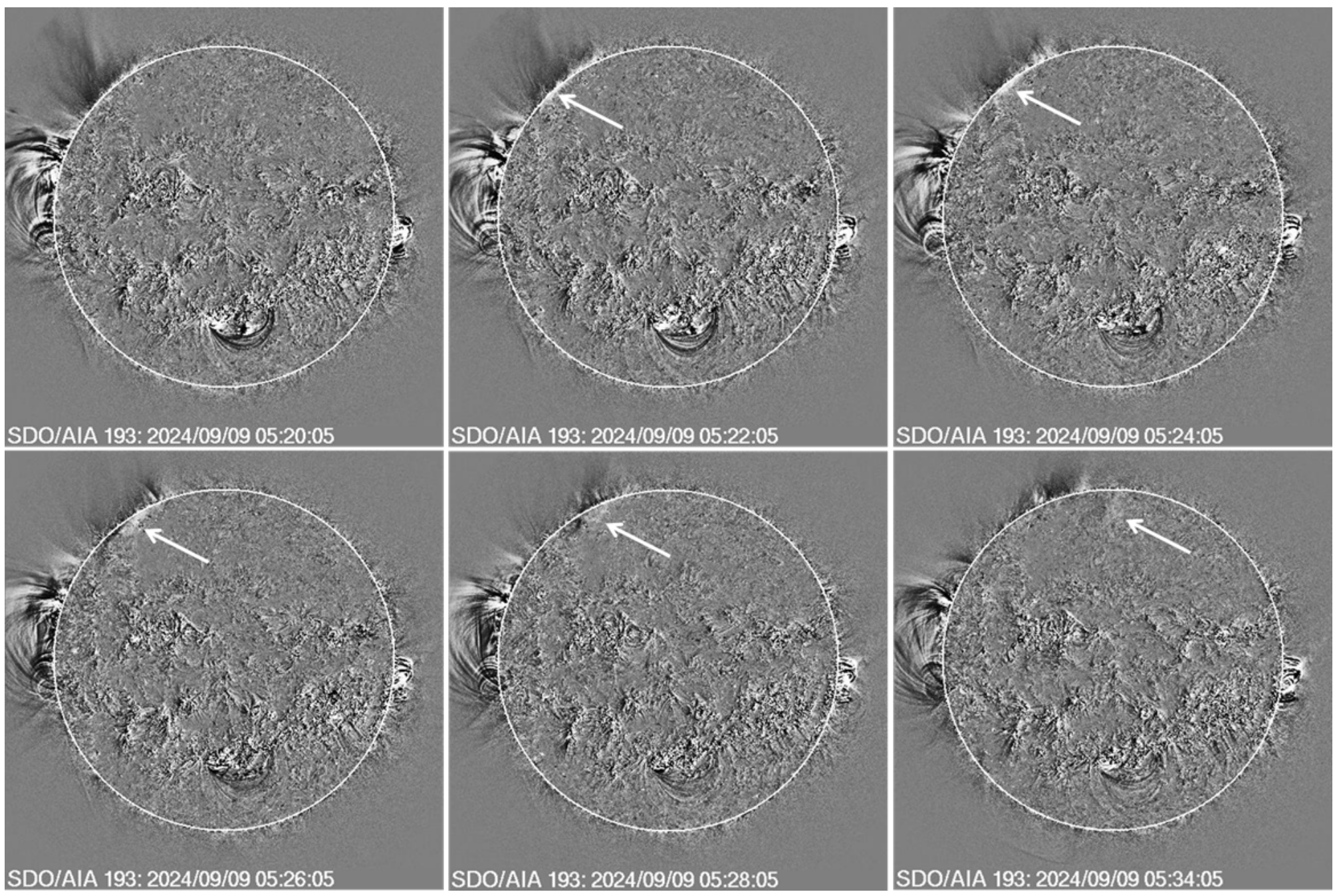

**Figure 9.** A series of SDO/AIA images showing the northward propagation of the EUV disturbance behind the east limb and its crossing to the frontside at 05:22 UT. The arrows point to the disturbance on the disk. A javascript movie of the disturbance can be viewed at: https://cdaw.gsfc.nasa.gov/movie/make_javamovie.php?date=20240909&img1=sdo_a193rdf.

**Table 2**. GCS parameters of the 2024 September 9 flux rope

| Ref Frame | UT | FR Direction | $R_{tip}$[1] (Rs) | θ (º) | κ | α (º) | $W_E$ (º) | $W_F$ (º) |
|---|---|---|---|---|---|---|---|---|
| EUVI | 05:03 | S12E130 | 1.42 | 37 | 0.11 | 6 | 12.6 | 24.6 |
| COR1 | 05:36 | N13E130 | 6.08 | 59 | 0.51 | 33 | 61.3 | 127.3 |
| COR2 | 06:38 | N13E130 | 16.5 | 59 | 0.58 | 53 | 70.9 | 176.9 |
| C3 | 06:54 | N13E130 | 18.8 | 59 | 0.58 | 53 | 70.9 | 176.9 |

[1]$R_{tip}$ = h + Ro, the leading-edge height of the flux rope.

### 2.5.2 Flux Rope Fitting

The coronagraph and EUV observations presented above provide a qualitative description of the event. These observations can be quantitatively combined to get the CME kinematics without projection effects. We fit a flux rope (FR) and its shock dome using the GCS model making use of near-simultaneous images in EUV and white light: SECCHI's EUVI, COR1, COR2, SDO's AIA, and LASCO's C2 and C3. The three parameters defining the FR geometry are the half-angle (δ) of the conical leg, the height of the leg (h), and the angle (α) between the axis of the cone and the axis of the FR apex, i.e. the half of the leg-separation angle. The FR geometry is defined using the parameter κ defined as the ratio of the FR radius (Ro) and the FR height h. The parameter κ = Ro/h sets the rate of expansion of the FR relative to the height, so that the FR structure expands self-similarly. Also, κ is related to the leg half angle by κ = sinδ. For comparing with observations, $R_{tip}$ = h + Ro, the leading edge of the flux rope is used. The edge-on ($W_E$) and face-on ($W_F$) widths of the FR are given by $W_E = 2\delta$ and $W_F = 2(\delta+\alpha)$. The heliocentric axis of the FR apex also defines the radial propagation direction of the FR, determined by the heliographic latitude (λ) and longitude (φ) of the intersection point of the axis and the solar surface. The tilt angle (θ) of the FR is defined as the angle between the heliographic east-west direction and the line connecting the FR legs. A positive (negative) θ indicates that the western leg of the FR has more northern (southern) latitude relative to that of the eastern leg. θ is measured from the west in the counterclockwise direction.

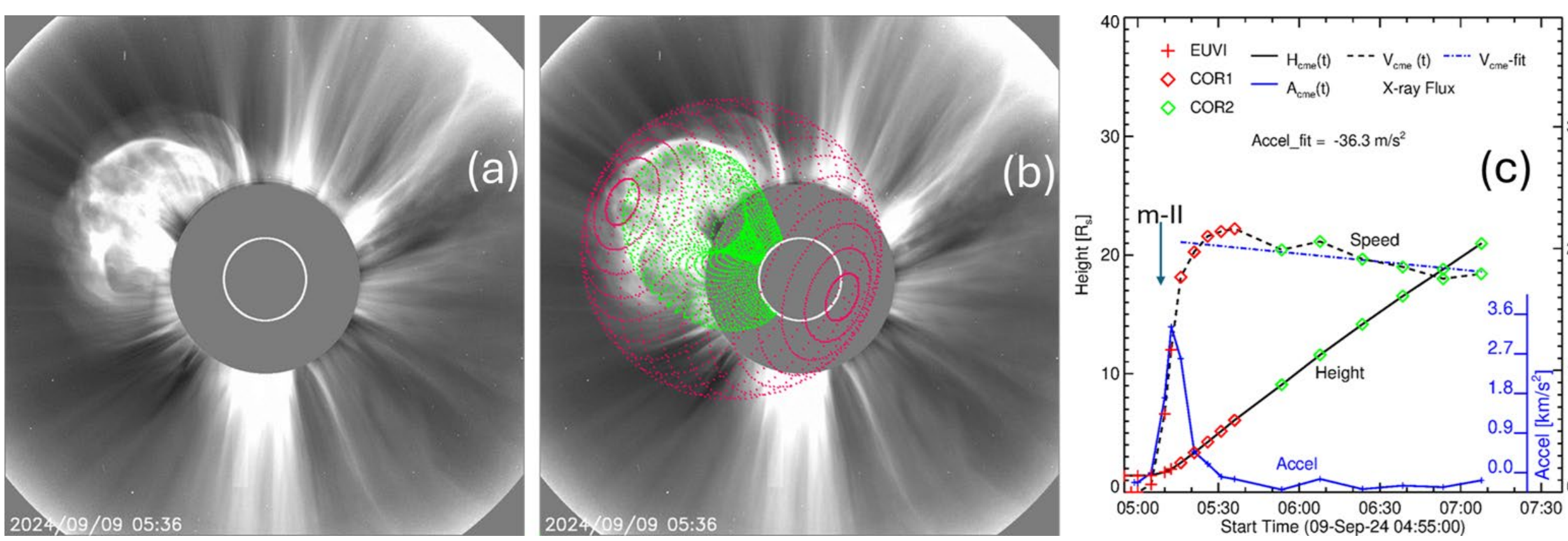

**Figure 10**. A LASCO/C2 image at 05:36 UT (a) with the GCS flux rope (green) and shock (red) superposed (b). The height-time, speed-time, and acceleration-time plots of the flux rope leading edge (c). At 05:12 UT the acceleration peaked with a value of 3.54 km $s^{-2}$. The peak speed (2162 km $s^{-1}$) was attained at 05:36 UT (same time as the images in (a) and (b)).

Figure 10 shows the fitted FR and shock superposed on a LASCO/C2 image at 05:36 UT. Table 2 shows the GCS parameters of the FR. When the FR attained its peak speed, it was at a height of 6.08 Rs. The initial direction of the flux rope is S12E130, which is close to the flare normal direction (S13E131). However, due to the northward deflection, the direction becomes N13E130, indicating a northward deflection by ~26º. The deflection is mostly in the latitudinal direction with little change in the longitudinal direction. By the time the FR reached a heliocentric distance of ~6 Rs, the deflection stops, and the FR direction remains at N13E130 even at a distance of 16 Rs. The height-time plot in Fig. 11c is obtained by tracking the tip of the flux rope. Clearly the FR accelerates impulsively attaining a peak of ~ 3.54 km $s^{-2}$ at 05:12 UT. Accordingly, the FR speed rapidly increases, reaching a peak value of ~2162 km/s at 05:36 UT, when the initial acceleration subsides followed by a slow deceleration (~36 m $s^{-2}$) due to drag. The FR speed decreases below 2000 km $s^{-1}$ around 06:20 UT, when it reached a height of ~13.5 Rs. The average speed in the coronagraph FOV (doing a linear fit excluding EUV data points) is 1936 km $s^{-1}$. Recall that the sky-plane speed of the CME LE obtained from LASCO is 1522 km $s^{-1}$. Since the source is ~41° behind the limb, the deprojected speed becomes 2017 km $s^{-1}$, consistent with the measurement using the FR fitting. Thus, the CME kinematics based on GCS fitting and simple deprojection consistently indicate that the CME is very energetic, similar to CMEs that produce GLE events and SGRE events.

The edge-on and face-on widths of the FR rapidly increased by a factor of ~5 until the FR reached its maximum speed at 5:36 UT, indicating a rapid expansion. After this time, the expansion significantly reduced. We also see that the tilt angle increased from 37° to 59° during the rapid expansion phase and then stabilized. Thus, the FR ended up having a high inclination. This is consistent with the fact that we see only the EUV wave around this time (see Fig. 10) because with high inclination, the FR still remains behind the limb but close enough that the accelerated protons precipitate to the frontside in the sheath region. Based on the face-on width, we estimate that the flux rope is at least 7° behind the limb.

### 3. Discussion

An ideal SGRE event under the shock paradigm meets several characteristics such as 1. Ultrafast halo CME (~2162 km $s^{-1}$ peak; 2017 km $s^{-1}$ average) similar to those that accelerate SEPs with ground level enhancement (GLE); 2. high initial acceleration (3.54 km $s^{-2}$ ), typical of CMEs resulting in GLEs; 3. hard-spectrum SEP Event (spectral index ~2.29); 4. IP type II burst extending below 1 MHz; 5. SGRE duration similar to the IP type II duration; 6. intense X-ray flare (~X3.3). The 2024 September 9 event satisfies all these characteristics and hence is consistent with a shock sheath surrounding the CME flux rope guiding the nose-accelerated protons to the frontside of the Sun to produce the observed SGRE.

**Table 3**. Backside eruptions that resulted in a Fermi/LAT SGRE event

| Date | Peak flux ($10^{-5}$ ph $cm^{-2}$ $s^{-1}$) | Sigma[1] | Duration (hours) | Fluence ($cm^{-2}$) | Source location | Flare size | CME speed[e] (km $s^{-1}$) |
|---|---|---|---|---|---|---|---|
| 2013 Oct 11 | 12.2 | 32.46 | 1.51 | 0.24 | N21E103 | M4.9[a] | 1208 H |
| 2014 | 1.31 | 2.11 | 1.10 | 0.01 | S15W112 | X3.5[a,b] | 1431 H |
| Jan 06 | | | | | | | |
| 2014 Sep 01 | 317.0 | 58.12 | 3.92±0.76 | 7.92 | N14E127 | X2.4[a] | 2017 H |
| 2021 Jul 17 | 1.02 | 4.77 | 1.48±1.58 | 0.01 | S20E140 | M5.0[b] | 1374 H |
| 2021 Sep 17 | 1.98 | 8.20 | 1.29± 1.62 | 0.04 | S30E100 | X1.9[c] | 1370 PH |
| 2022 Sep 29 | 0.69 | 2.04 | 0.78±0.61 | 0.007 | N26E106 | X2.0[d] | 416 PH[f] |
| 2024 Feb 14 | 1.94 | 7.69 | 1.93± 0.79 | 0.06 | S36W160 | ---- | 2456 H |
| 2024 Sep 09 | 0.81 | 3.38 | 1.75±0.79 | 0.02 | S13E131 | X3.3[d] | 2017 H |

[1]Number of (effective) sigmas above mean. [a]From Ajello et al. (2021); [b]Pesce-Rollins et al. (2022); [c]Estimated based on the number of saturated EUVI pixels (Yashiro et al. 2024); [d]Estimated based on the SolO/STIX observations; [e]H and PH denote halo and partial halo CMEs; The CME was impulsively accelerated to a speed of ~1900 km/s early on and decelerated to 416 km/s in the LASCO FOV.

Pesce-Rolllins et al. (2022) recently compiled and compared four BTL events detected by Fermi/LAT until 2021 September 17 to show that the same agent is responsible for the EUV waves and SEPs, viz., the shock. We have listed these BTL events in Table 3 along with additional BTL events that occurred on 2014 January 6, 2022 September 29, 2024 February 14, and 2024 September 9. Thus, the total number of BTL events now stands at 8 over the entire SC 24 (3 events) and up to the maximum of SC 25 (5 events). The 2022 September 29 gamma-ray event was extensively discussed in Pesce-Rollins et al. (2024). The gamma-ray duration in this event has been computed from the GOES soft X-ray peak at 12:01 UT to 12:46:30 UT, which is the mid-time between the end of the first Sun exposure at 12:10 UT and the next exposure beginning at 13:23 UT. We have listed the peak flux, duration, and fluence of the SGRE events. We have also given the heliographic coordinates of the eruption location, estimated X-ray flare size, and deprojected CME speed stating whether the CME is a halo. We see that the peak fluxes and fluences are generally low except for the 2014 September 1 event, which remains the most intense event. Gopalswamy et al. (2019) reported the average fluence of >3 h SGRE events to be 0.49 $cm^{-2}$. All events in Table 3 have fluences below this average, except for the 2014 September 1 event. For example, the 2024 September 9 event is very weak, with a peak flux and fluence that are lower by a factor of ~400 relative to the 2014 September 1 event. Based on a simple model of the extended gamma-ray source, Gopalswamy et al. (2021) showed that the

>100 MeV flux/fluence should be much larger than what is observed in BTL events because most of the source is hidden behind the limb. Assuming a source radius of 40º, they estimated that the observed flux in the 2014 September 1 event might have been reduced by a factor of ~560. For a similar source size in the 2024 September 9 event, the flux/fluence reduction is even more (by a factor of ~3300) because the source is farther from the limb by ~4º, and the reduction is an extremely steep function of the longitudinal distance behind the limb (Gopalswamy et al. 2021, their Fig. 2). With such a huge correction factor, the 2024 September 9 event is smaller than that of the 2014 September 1 event only by a factor of ~65. One expects a similar correction factor in the case of the 2021 July 17 event because it is ~50º behind the limb. It must be noted that in the 2024 September 9 event, the Sun exposure started ~26 min after the expected start of the on-disk gamma-rays. Given the limited coverage of the gamma-ray event, the observations do not cover the time in which the flux peaked.

The source locations of the backside events are obtained either by tracking the region from front to back or from SolO/STIX flare locations. We see that the source longitudes vary from W160 to E140 (i.e., 70º behind the west limb to 50º behind the east limb). The flare sizes range from M4.9 to X3.5. The 2024 September 9 event has the second largest flare size. SolO/STIX imaged the HXR and SXR flare structures in the 2021 July 17 and 2024 September 9 eruptions. Unlike the HXR ribbon structure in the 2024 September event, the 2021 July event showed just two HXR foot point sources straddled by the SXR structure.

All CMEs in Table 3 are halos except for the 2021 September 17 CME, which is a partial halo with a width of 264º. The speeds of halos are all deprojected using a cone model and are listed in the halo CME list (https://cdaw.gsfc.nasa.gov/CME_list/halo/halo.html, Gopalswamy et al. 2010). The CME speeds range from 1208 km $s^{-1}$ to 2456 km $s^{-1}$. The 2024 September CME has the same speed as the 2014 September event. The 2024 February 14 CME was the fastest (2456 km $s^{-1}$) in the table.

There are only two events in Table 3 that had high-energy SEPs observed in space: our event (see Fig. 6) and the 2014 January 6 SGRE that had >700 MeV protons observed by GOES and also GeV particles observed by neutron monitors (Thakur et al. 2014). This event also showed northward deflection but only by ~9º (Gopalswamy et al. 2014) as compared to 26º in the 2024 September 9 event. The fluence spectral index (2.54) of the 2014 January event is slightly larger than that (2.29) of the 2024 September event.

There is one fact that stands out in Table 3 is that all BTL SGRE events are behind the east limb except for the 2014 January 6 and 2024 February 14 events. The 2014 January 6 event had the most intense flare in Table 3 and was only 22º behind the west limb. Yet, the SGRE peak flux, fluence, and duration are all modest. In the case of the 2024 February 14 event, the source is at S36W160. We identified the source location by tracking NOAA AR 13575, which produced the previous SGRE event on 2024 February 9 when it was at the west limb. In longitude, this event is farthest from the west limb (~70º). However, the source latitude is at S36, which means, the south limb is only 44º away. This suggests that the protons seem to be precipitating to the frontside over the pole.

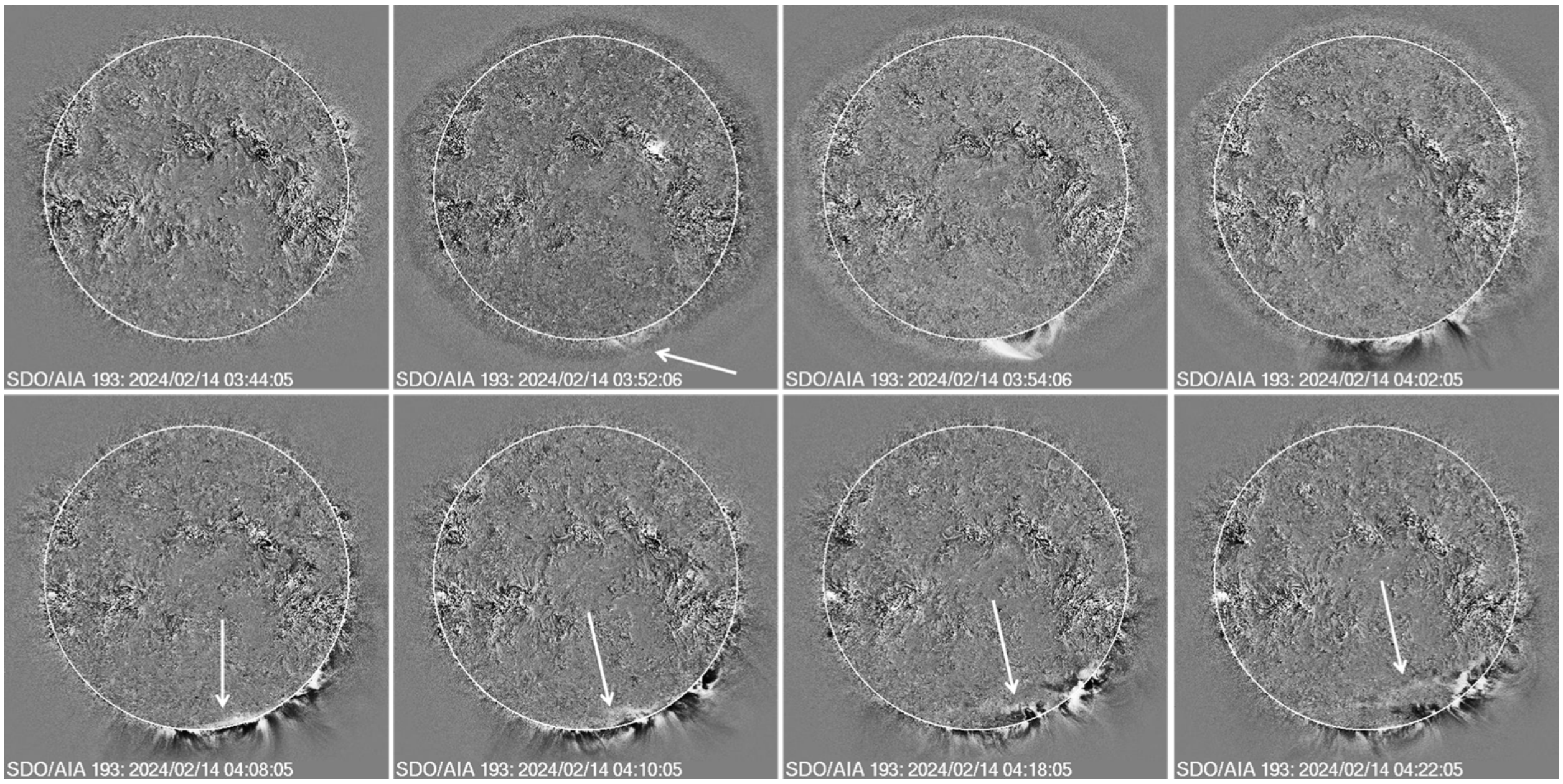

**Figure 11**. The 2024 February 14 EUV disturbance from SDO/AIA crossing the limb near the south pole. The arrow in the 03:52 UT frame shows the first appearance of the disturbance above the limb. The arrows in the bottom panel point to the disturbance propagating on the disk.

Figure 11 shows the evolution of the EUV disturbance from the backside eruption. The EUV manifestation of the CME first appears above the south limb at 03:52 UT. In the next two minutes, the CME reaches the edge of the SDO/AIA FOV. The disturbance can be seen spreading along the limb until 04:02 UT. By 04:06 UT, the disturbance reached the south limb from behind and crossed over at 04:08 UT as can be seen in Fig. 11. The SGRE was observed around this time. The timing of the limb crossing and the start of on-disk SGRE are very close, much better than the situation in the 2024 September 9 event. Crossing of the EUV wave over the poles confirms the possibility of frontside precipitation of protons over the pole. Recall that we showed the wave crossing the east limb at high latitudes in Fig. 9 due to the northward deflection of the CME. In the case of the 2024 February 14 eruption, the limb crossing is in the latitudinal direction owing to the location of the active region at a latitude of S36. Details on the SGRE event including the localization of the gamma-ray source  will be reported elsewhere. We point out that the angular distance of the eruption region to the limb is similar to the 2024 September 9 event, although in the latitudinal direction.

Returning to the east-west asymmetry, it is quite intriguing that the BTL events occur preferentially from behind the east limb. The backside extreme eruption of  2012 July 23 originated at a longitudinal distance similar to that of the 2021 July 17 event but behind the west limb (S17W141). GCS fit to that event yielded a flux rope location N05W135 (Gopalswamy et al. 2016), similar to the 2024 September 9 event in terms of the angular distance behind the limb. Yet, there are no gamma-rays from this event (there was LAT coverage) in spite of the hard-spectrum SEP event observed by STA. There was a small GLE event (#76, https://gle.oulu.fi/#/) observed on 2024 November 21 at ~18:05 UT in association with and X3.0 flare (estimated) and the 18:12 UT halo CME (speed ~1626 km $s^{-1}$) from BTL at S10W118 (https://cdaw.gsfc.nasa.gov/movie/make_javamovie.php?stime=20241121_1651&etime=20241121_2040&img1=lasc2rdf&title=20241121.181205.p295g;V=1436km/s). There seems to be a Fermi/LAT data gap between November 19 and December 9 in 2024, so we do not know if this GLE was associated with an SGRE event. If this GLE turns out to be an SGRE event, then the asymmetry might reduce. As of this writing, the east-west asymmetry stands at 3:1. The asymmetry is even larger if we do not include the 2024 February 14 event in which the wave is over the south pole rather than from behind the west limb. The east-west asymmetry needs to be investigated further to see if this has anything to do with the asymmetric distribution of magnetic field lines on the eastern and western side of the shock sheath and the inclination of the CME flux rope.

## 4. Conclusions

The 2024 September 9 SGRE event originated from an eruption that occurred ~40º behind the east limb, very similar to the 2014 September 1 event in terms of eruption longitude. However, the 2024 event had a much weaker >100 MeV peak flux and fluence by more than two orders of magnitude. The 2024 CME was driving a shock that produced a type II radio burst in the metric and decameter-hectometric wavelength domains. The shock also accelerated SEPs to very high energies, up to ~1 GeV. The eruption involved an intense soft X-ray flare, which was estimated to have a GOES-equivalent flare size of ~X3.3. Forward modeling of the CME flux rope and shock revealed that the flux rope was deflected northward by 26º with a significant rotation. The flux rope was an ultrafast CME (2162 km $s^{-1}$) that had a high initial acceleration (3.54 km $s^{-2}$). These characteristics are typical of a front-side Fermi/LAT SGRE events, although most SGRE events are associated with M- and X-class soft X-ray flares. An EUV wave associated with the flux rope crossed the limb, and was observed on the disk supporting the scenario that the shock sheath extended to the frontside of the Sun where the high-energy protons precipitated to produce the observed gamma-ray emission. Although the temporal sequence (on-disk SGRE followed by EUV wave limb crossing) is correct, the timing is less than perfect due to a large Fermi/LAT data gap. The timing is much better in most other BTL events: 2013 October 11, 2014 September 1, 2021 July 17, 2021 September 17, and 2024 February 14 BTL events. The main conclusions of this study are as follows:

1. The 2024 September 9 BTL SGRE is a relatively small event but had all the signatures of high-energy eruptions that cause SGRE.

2. The eruption was associated with an energetic halo CME that impulsively accelerated (3.54 km $s^{-2}$) to attain speeds exceeding 2000 km $s^{-1}$.

3. The eruption involved an X3.3 flare that had hard X-ray ribbons and soft X-ray thermal structure.

4. The event had metric and IP type II bursts indicating shock formation early in the event when the CME leading edge was at a height of ~1.5 Rs, which is typical of high-energy SEP events such as GLE events. The metric type II burst continues into the DH domain, most likely indicating flank emission.

5. A second component of the type II bursts starts at frequencies below 10 MHz, which seems to be from the shock nose.

6. Type III burst activity is delayed with respect to the impulsive phase of the flare and the onset of type II activity indicating that the underlying electrons might be accelerated by the shock rather than by the flare reconnection.

7. Forward modeling indicates that the CME flux rope was very large (full width ~176º), deflected northwards, and rotated counterclockwise by 22º.

8. Like in most BTL SGRE events, EUV waves crossed from behind the limb onto the frontside consistent with energetic proton precipitation to the frontside behind the shock to produce the gamma-rays as observed by Fermi/LAT around the time of limb-crossing. The timing is less than desirable mainly because of a 26 min Fermi/LAT data gap.

9. The flux rope drove a shock visible in white light and resulted in a large hard-spectrum SEP event that had protons up to GeV energy indicating there were the required >300 MeV particles to produce gamma-rays via the pion decay mechanism.

10. Comparing with all the BTL events observed by Fermi, we found a 3:1 asymmetry between events behind the east and west limbs, suggesting a possible preferred proton propagation on the western side of CME flux ropes.

**Data Availability**

The gamma-ray data used in this manuscript is publicly available at https://hesperia.gsfc.nasa.gov/fermi_solar/. The CME, SEP, and radio burst data are publicly available at the CDAW Data Center (https://cdaw/gsfc.nasa.gov) and https://parker.gsfc.nasa.gov/crocs.html. The ground-based radio data are available at https://e-callisto.org. Solar Orbiter data are available at https://espada.uah.es/epd/EPD_data.php and https://datacenter.stix.i4ds.net/.

**Funding**

NG is supported by Goddard Space Flight Center, LWS-HSSO, STEREO Project, and NASA Science Mission Directorate, 2024 HISFM. PM is supported by Goddard Space Flight Center, PHaSER-670.089 and US National Science Foundation, AGS-2043131. SA and HX are supported by Goddard Space Flight Center, PHaSER-670.089 and US National Science Foundation, AGS-2228967. SDB is supported by NASA contract NNN06AA01C. RFWS and PK are supported by Deutsches Zentrum für Luft- und Raumfahrt, 50OT2002. SK is supported by Swiss PRODEX grant for STIX.

**Acknowledgements**

The Fermi mission is a joint venture of NASA, the United States Department of Energy, and government agencies in France, Germany, Italy, Japan, and Sweden. SOHO is a project of international collaboration between ESA and NASA. STEREO is the third mission in NASA's Solar Terrestrial Probes (STP) program. Parker was designed, built, and is operated by the Johns Hopkins Applied Physics Laboratory as part of NASA's LWS program. The PSP/FIELDS experiment was developed and is operated under NASA contract NNN06AA01C. Solar Orbiter is a space mission of international collaboration between ESA and NASA, operated by ESA. SolO/EPD work at CAU is supported by the German Federal Ministry for Economic Affairs and Energy and the German Space Agency (Deutsches Zentrum für Luft- und Raumfahrt, e.V., (DLR)), grant number 50OT2002. The SolO/STIX instrument is an international collaboration between Switzerland, Poland, France, Czech Republic, Germany, Austria, Ireland, and Italy. SDO is part of NASA's LWS program. The Global Geospace Science (GGS) Wind is a NASA science spacecraft designed to study radio waves and plasma that occur in the solar wind and in the Earth's magnetosphere. Geostationary Operational Environmental Satellites (GOES) is a collaborative NOAA and NASA program providing continuous imagery and data. CALLISTO is an instrument network of the International Space Weather Initiative (ISWI). The GLE database is hosted and managed by the Oulu Cosmic Ray Station of the University of Oulu, Finland.